\documentstyle[iopconf1,epsfig]{article}
\newcommand\place[4]{
  \begin{center}
      \mbox{\psfig{file=#2,width=#1\textwidth}}
      \caption[]{#4}
      \label{#3}
  \end{center}}
\begin{document}
\title{New physics at LEP2~\footnote{To appear in the proceedings of ``Beyond
     the Desert'', Castle Ringberg, Tegernsee, Germany, 8-14 June 1997.}}
\author{Ramon Miquel~\footnote{E-mail address: miquel@ecm.ub.es}}
\affil{Dept.~ECM and IFAE \\
       Univ.~de Barcelona, Facultat de F\'{\i}sica \\
       Diagonal 647, E-08028 Barcelona (Spain)}
\beginabstract
Searches for new physics in the first year of running of LEP2, at energies
of 161 and 172~GeV, are summarized. After a short review of WW results and
their implications on new physics,
searches for the Higgs
boson and SUSY particles, analyses of four-jet final states and constraints on
possible explanations for the HERA high $Q^2$ anomaly are discussed in turn.
\endabstract
\section{Introduction}
The second phase of LEP, LEP2, designed to run at energies above the W-pair 
production threshold started operation in 1996, with two short runs, one at a
center-of-mass energy of 161~GeV and another at about 172~GeV. The four LEP
experiments, ALEPH, DELPHI, L3 and OPAL, collected about 10~pb$^{-1}$ of 
luminosity each in each of the two runs.

The main purpose of the 161~GeV run was to measure the W-pair production cross
section very close to threshold. This cross section is very sensitive to the
value of the W mass and it provides a measurement with good
precision. Searches for new particles were also performed at this energy, but
they have already been superseded by the searches done at 172~GeV soon after.

In the 172~GeV run, the mass of the W gauge boson was determined directly,
through the measurement of the invariant mass of its decay products. Also
a first look at the structure of the trilinear gauge-boson vertices was
attempted. A short summary of W physics in the first year of LEP2 and its
implications on limits for new physics can be found in section~\ref{sect:WW}.

Results from the searches in the 172~GeV run for Higgs bosons (both from the
Standard Model and from Supersymmetry models) and for supersymmetric 
particles are summarized in sections~\ref{sect:Higgs} and \ref{sect:SUSY}, 
respectively.

In the 1995 LEP run at a center-of-mass energy about 130~GeV, the ALEPH 
collaboration reported an excess of four-jet events with sum of jet-pair
invariant masses close to 105~GeV~\cite{ref:aleph_133}. The analyses of ALEPH
and the other LEP collaborations on this topic both in the 161~GeV run
and in the 172~GeV run are discussed in section~\ref{sect:4-jet}.

The reports of the H1 and ZEUS collaborations at HERA hinting at possible new
physics in electron-quark interactions~\cite{ref:HERA} has prompted several
analyses of electron-positron anihilation into quarks that are sensitive to
some types of new physics that could explain the HERA events. 
Section~\ref{sect:HERA} summarizes the status after analysing the 1996 data.

Finally, section~\ref{sect:summary} contains a summary of the talk. It should
be noted that, unless specified otherwise, all results are to be understood
as still preliminary.
\section{W physics}
\label{sect:WW}
The precise determination of the W mass serves as a stringent test of the 
Standard Model, since it can be predicted from the known values of the other
parameters of the Standard Model. The only large
uncertainty comes from the lack
of knowledge of the Higgs boson mass, and therefore, measuring $M_W$ precisely
enough, one can get information on the value of $M_H$.

The measurement of the W cross section close to the W-pair production threshold
provides a clean and precise way to determine the W mass, since the phase-space
factors appearing in the cross section expression 
depend strongly on the W mass.

All four LEP experiments have already published their final results on the WW
cross section production at 161~GeV~\cite{ref:WW_161}. The combined value for
the WW cross section is~\cite{ref:LEPEWWG}
$$
\sigma_{WW}(161.3\,\mbox{\rm GeV}) = 
\left( 3.69 \pm 0.45 \right)\mbox{\rm  pb}\, .
$$
The error is dominated by statistics.
From the total WW cross section at threshold, the W mass is obtained by
using a Standard Model calculation that relates the WW cross section near
threshold to its mass. The resulting W mass is:
$$
M_W (\mbox{\rm 161 run}) = 
\left(80.40 ^{+0.22}_{-0.21} \pm 0.03 \right) \mbox{\rm  GeV}\, ,
$$
where the last error reflects the current uncertainty on the LEP beam energy.

Above the WW threshold, the most efficient way of determining the W mass is by
measuring the invariant mass distribution of its decay products, either a 
charged lepton and a neutrino or a pair of jets. 
The detector resolution being far too poor to obtain a reasonably narrow
invariant mass distribution, energy and momentum conservation have to be 
imposed in all channels to improve the resolution.

The preliminary determinations of the four experiments agree very well with
each other and the combined value reads~\cite{ref:LEPEWWG}
$$
M_W (\mbox{\rm 172 run}) = 
\left(80.37\pm 0.18_{exp}\pm 0.05_{theo} \pm 0.03_{beam}\right)
\mbox{\rm  GeV}\, .
$$
The dominant error includes statistical and purely experimental
errors, the second one is the estimate of the uncertainty due to soft QCD
effects and the third comes from the beam energy uncertainty.
\begin{figure}[htb]
\begin{center}
\place{1.0}{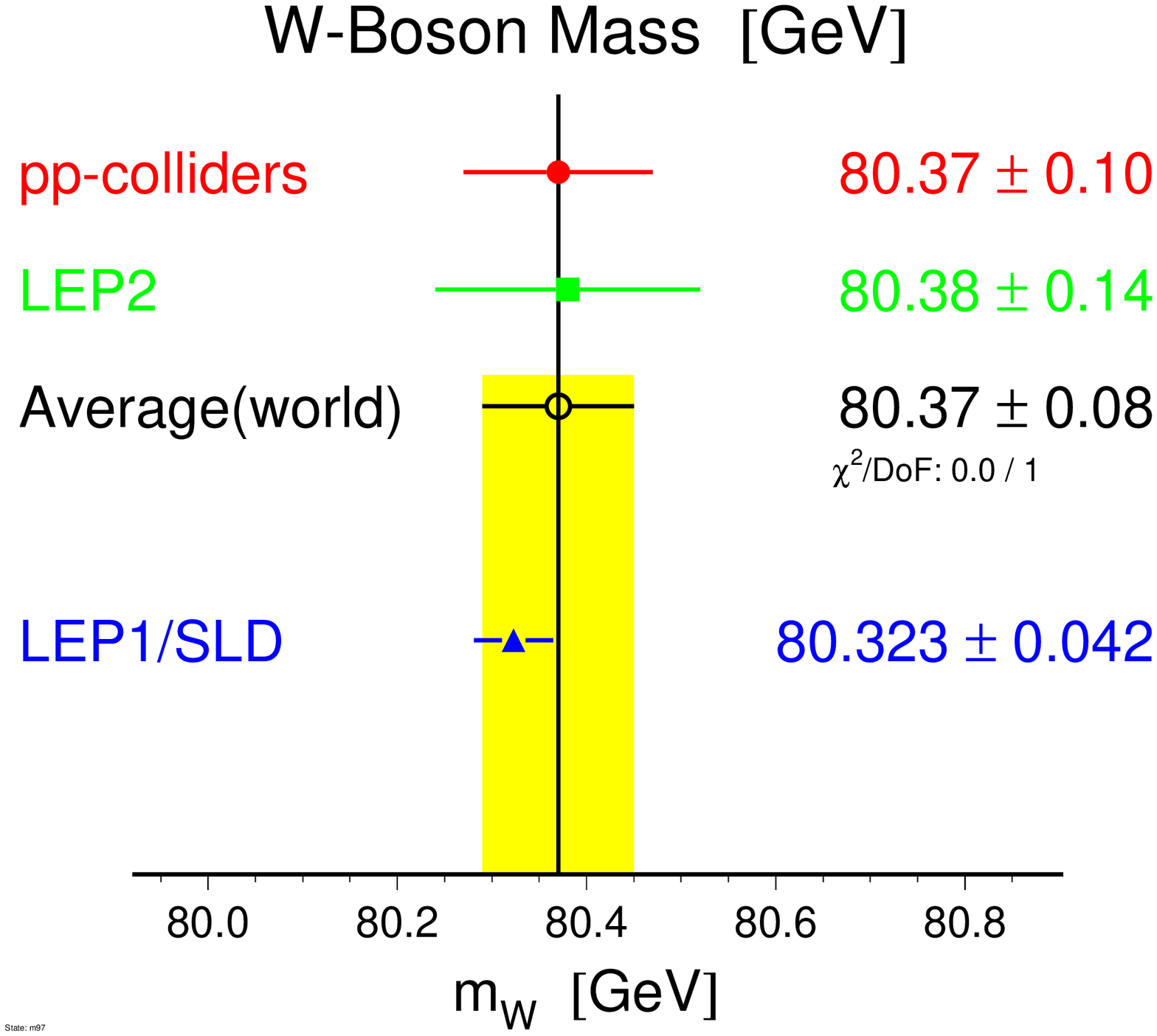}{fig:mwcompa}
{Comparison of the direct W mass measurements at the Tevatron and LEP2 and 
the indirect determination using LEP1 and SLD data.}
\end{center}
\end{figure}

The two numbers can be combined~\cite{ref:LEPEWWG} to give the 
preliminary W mass result from the first year of running of LEP2:
$$
M_W = \left(80.38 \pm 0.14\right)\mbox{\rm  GeV}\, .
$$
This value is compared in fig.~\ref{fig:mwcompa} with the other measurements
of $M_W$, both direct at the Tevatron and indirect at LEP1 and SLD. 

The agreement of all
measurements is perfect. The current LEP2 error is already quite good 
but still much larger than the 40~MeV uncertainty
in the indirect measurement. To get a direct determination of $M_W$ with this
sort of accuracy and to compare it against the indirect measurement, which
assumes the Minimal Standard Model, is the next challenge for both LEP2 and
the Tevatron. 

The other important topic of WW physics at LEP2 is the search for possible
anomalous couplings between a pair of Ws and a Z or a photon. However, these
studies require the highest possible energy as well as substantial amount of
integrated luminosity. The results available so far, 
which include only the semileptonic WW decays after the 172~GeV run,
only improve marginally on the limits obtained previously at the Tevatron.
\section{Higgs search}
\label{sect:Higgs}
The main production process for the Standard Model Higgs at LEP2 is its
production associated with an on-shell Z. The cross section goes down to a
fraction of a picobarn once the Higgs mass reaches about $\sqrt s - 100$~GeV.
This is, more or less, the discovery limit for a fixed center-of-mass energy.
Since the LEP1 limit stands at about 65~GeV, it is clear that the 161~GeV run
was not useful for Standard Model Higgs search, while the 172~GeV run 
started to extend the search region.

The main decay channel for a Higgs boson with mass around 70~GeV is to a pair
of $b$ quarks. Therefore, according to the Z branching ratios,
the final state $ZH$ will consist 70\%
of the times of four jets, two of them $b$s; 20\%
of the times of two $b$ jets and missing energy; and 10\%
of the times of two $b$ jets and two charged leptons. The four-jet channel is
both the most aboundant and the most difficult to separate from the background.
Good b-tagging capabilities are mandatory and all the LEP experiments have
invested recently in new and more powerful silicon vertex detectors. Typical
efficiencies in the four-jet channel are around 30\%
with very low background contamination.
\begin{figure}[p]
\begin{center}
\place{1.0}{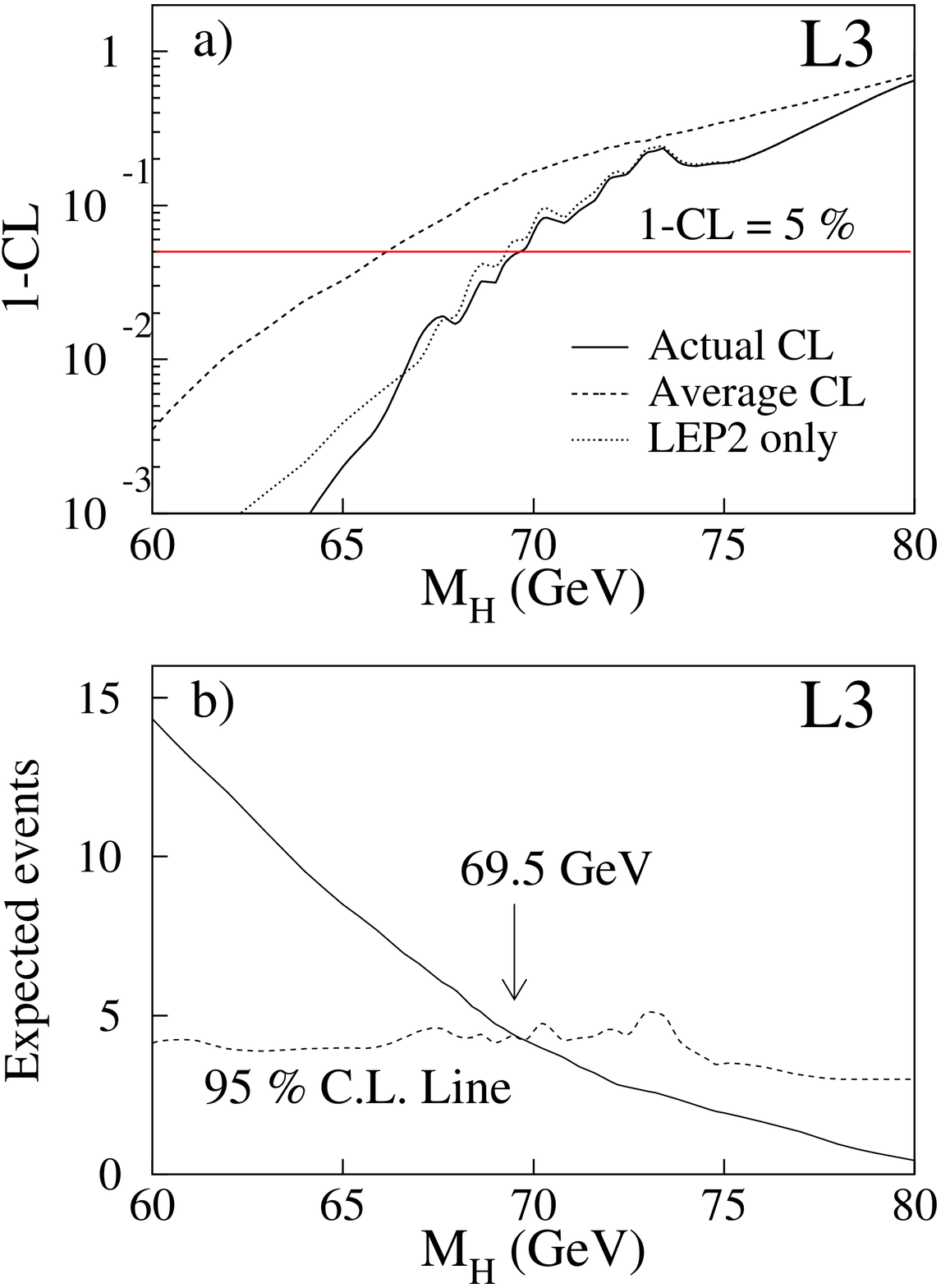}{fig:SMH_aleph}
{a) Exclusion confidence level as a function of the Higgs mass for L3.
b) Expected number of Standard Model Higgs events in L3 as a function of the
Higgs mass. The arrow
points at the mass excluded at the 95\%
confidence limit.}
\end{center}
\end{figure}
No experiment has found any evidence for an excess in any of the channels that
have been searched. Typical 95\%
confidence level
limits on the mass of the Standard Model Higgs boson
stand at around 70--71~GeV~\cite{ref:higgs}.
Figure~\ref{fig:SMH_aleph} shows the number
of Higgs bosons expected by L3 as a function of $M_H$, and the limit taking
into account the observed candidates (compatible with background expectations).

In all Supersymmetry models at least two Higgs doublets appear, giving rise 
to five physical states: two charged Higgses, two neutral CP-even Higges
($h,H$) and a neutral CP-odd Higgs ($A$). In most SUSY theories there is an 
upper limit to the mass of the lightest CP-even neutral Higgs,
$h$, which is around 150~GeV~\cite{ref:kane}.
Furthermore, in many models, its mass is below 100~GeV, suitable for its
search at LEP2.

The lightest CP-even Higgs can be produced in association with a Z, in a
process very similar to its Standard Model counterpart, or in association with
the CP-odd state, $A$. Both processes are complementary, in the sense that in
the regions of the SUSY parameter space in which one cross section is small the
other is large and viceversa, so that the 
overall rate of $h$ production remains
sizable in all regions with $M_h, M_A < 60-65$~GeV.
\begin{figure}
\begin{center}
\place{0.9}{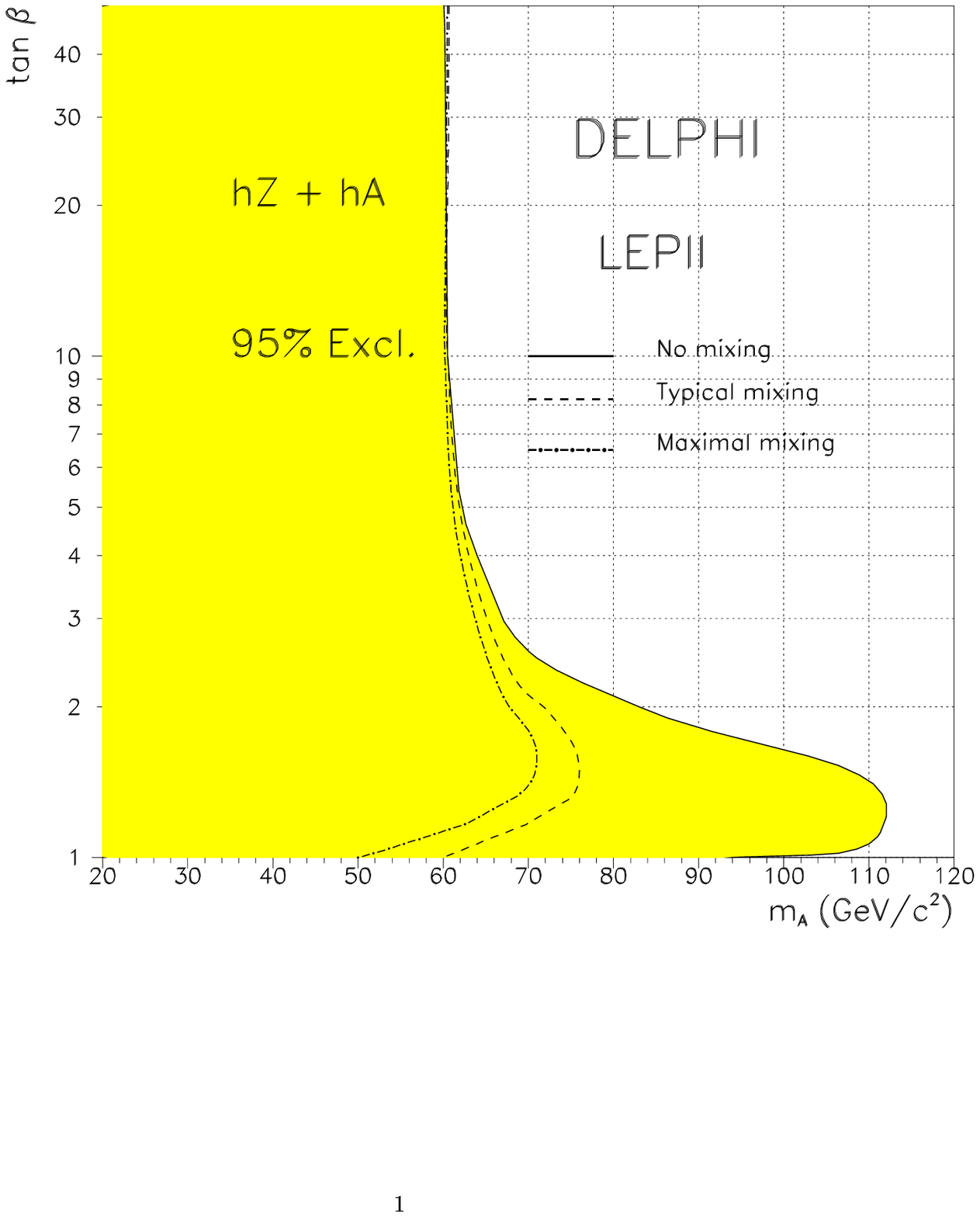}{fig:ma-tanb}
{Regions of the $M_A$--$\tan\beta$ region excluded by the DELPHI collaboration
under several assumptions.}
\end{center}
\end{figure}

Since all neutral Higgses decay predominantly to $b$ jets, the channel $hA$
involves identifying a four-jet event with four $b$s in the final state.
As in the case of the Standard Model Higgs, no excess has been 
found~\cite{ref:higgs} in any channel. Absolute 95\%
confidence level mass limits on both $h$ and $A$
have been set at about 62.5~GeV for all values of $\tan\beta\leq 1$, 
where $\tan\beta$ is the ratio of the vacuum expectation value of the
Higgs doublet that gives mass to up-type particles to that of the doublet
that gives mass to down-type particles. Figure~\ref{fig:ma-tanb} shows
the region in the plane $M_A-\tan\beta$ excluded by the DELPHI
collaboration.
\section{Supersymmetry searches}
\label{sect:SUSY}
Supersymmetry models not only predict more fundamental scalars. They predict
for every particle in the Standard Model another particle with spin differing 
by $\pm 1/2$,
the supersymmetric partner. 
The partners of the charged Higgses and Ws are generally called charginos, 
while the partners of the neutral Higgses and neutral 
electroweak vector bosons are
known as neutralinos. Charginos and neutralinos are expected to be the 
lightest SUSY particles. In particular, the lightest neutralino, $\chi$, 
is supposed to be the 
lightest supersymmetric particle (LSP) in most models with gravity-mediated
SUSY breaking.

Searches have been performed for chargino pairs ($\chi^+\chi^-$), neutralino
pairs ($\chi\chi'$), where $\chi'$ is the second lightest neutralino, 
scalar-lepton pairs ($\tilde{l}^+\tilde{l}^-$), scalar-top pairs
($\tilde{t}_1\bar{\tilde{t}}_1$), etc. If R-parity is conserved, then the LSP,
assumed to be $\chi$, is stable and does not interact in the detector, so that
the main experimental signature for the
production of SUSY particles is the presence of large missing energy in the
event. 

Several topologies involving jets and/or leptons and missing energy have been
searched for by the four collaborations with negative results. Therefore,
limits on the masses of the SUSY particles have been put. The
cross sections used to derive the limits on the masses and the masses 
themselves depend on many of the parameteres of the SUSY models, making 
difficult the
task of giving absolute limits.
\begin{figure}
\begin{center}
\place{1.0}{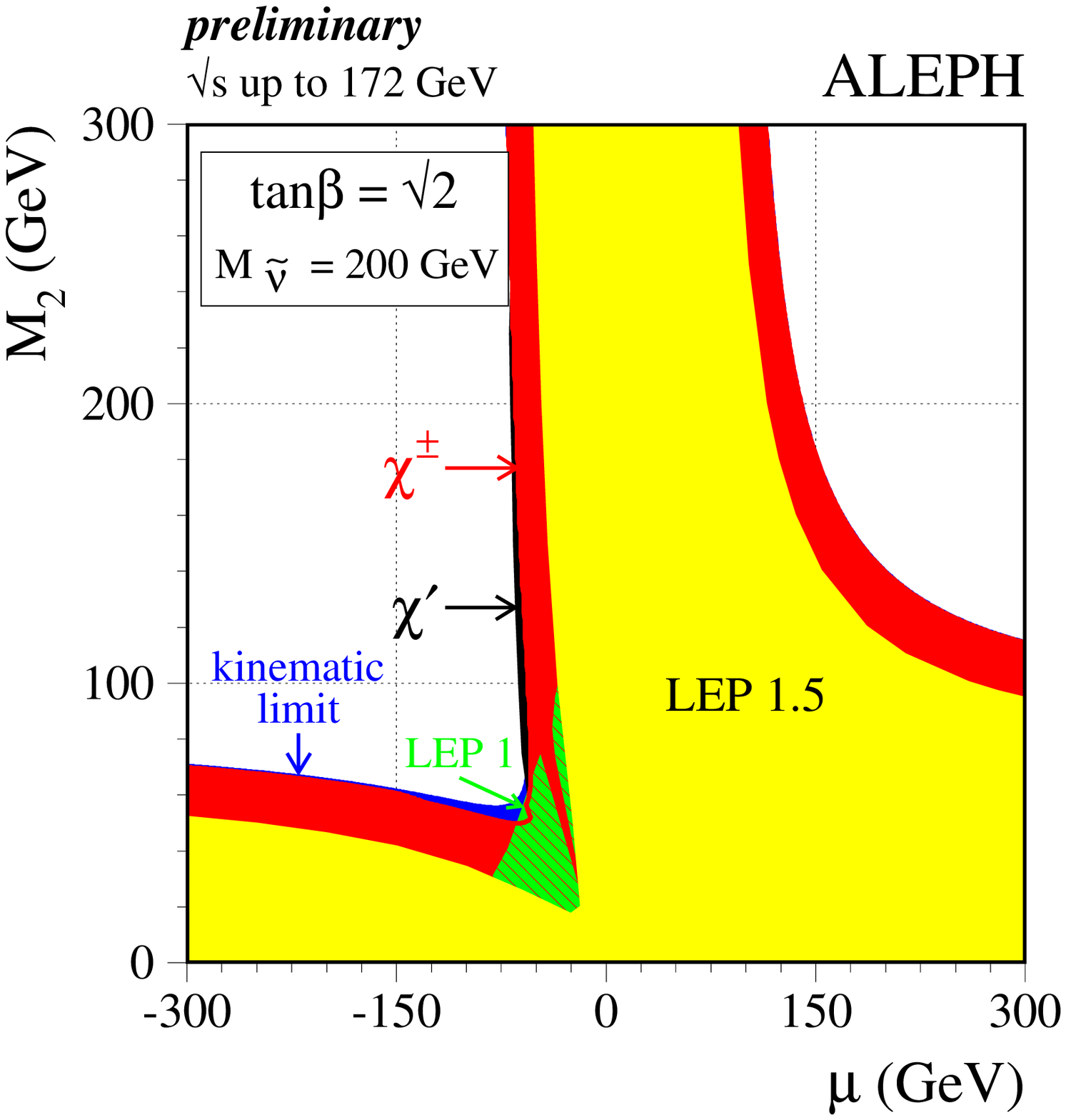}{fig:m2_mu}
{Regions of the $M_2$--$\mu$ region excluded by the ALEPH collaboration
for $\tan\beta=\sqrt{2}$ and scalar neutrino mass of 200~GeV.}
\end{center}
\end{figure}

In most models, one can set a limit on the mass of the lightest chargino close
to the kinematical limit, about 85~GeV. If, however, the scalar neutrino mass
is light (below 100~GeV), the limit degrades because the cross section for
chargino production decreases, due to the diagram with a t-channel scalar
neutrino exchange. The mass limit for the lightest neutralino is only
about 24~GeV, assuming heavy scalar leptons. The result of the combined 
searches for charginos and neutralinos is customarily displayed as excluded
areas in the $M_2-\mu$ plot, where $M_2$ and $\mu$ are gauge- and Higgs-mass
parameters appearing in the SUSY lagrangian. One such plot, showing the 
regions excluded by ALEPH, can be seen in fig.~\ref{fig:m2_mu}.
\begin{figure}
\begin{center}
\place{0.9}{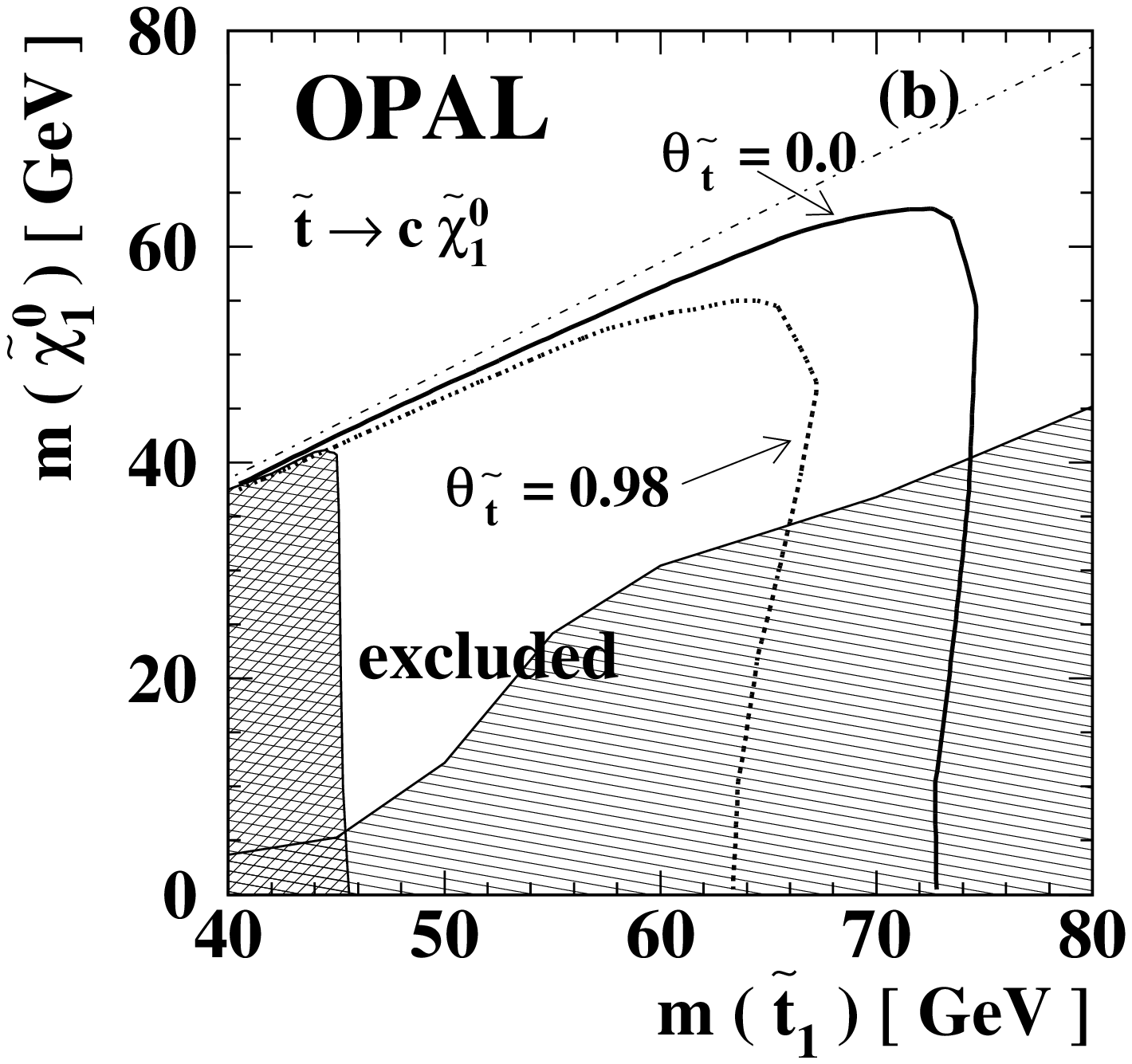}{fig:stop_opal}
{Regions in the $m_{\tilde{t}_1}-m_{\chi}$ plane excluded
by OPAL~\cite{ref:stop_opal} for several values of the stop mixing angle, 
assuming the decay goes as $\tilde{t}_1 \to c\chi$ 100\%
of the times. The cross-hatched area had already been excluded at LEP1, while
the single-hatched area has been excluded by D0~\cite{ref:stop_d0}.}
\end{center}
\end{figure}

Limits for scalar partners of leptons from 53~GeV (for $\tilde{\tau}$) to
75 GeV (for $\tilde{e}$) have been put. Assuming that the lightest scalar top
decays 100\%
of the times to $c\chi$,
the limit on its mass is about 65~GeV for the value of the stop mixing angle 
which results in the smallest cross section. For some other values, the limit
goes up to 73~GeV, as can be seen in fig.~\ref{fig:stop_opal} from 
OPAL~\cite{ref:stop_opal}.
More details on all these limits can be
found, for instance, in ref.~\cite{ref:fabio}.

In models in which supersymmetry is broken via gauge interactions, the 
gravitino (SUSY partner of the graviton) is the LSP. 
Then, if $\chi$ is the
next-to-lightest SUSY particle, the production of a pair $\chi\chi$ can 
result in a final state $\gamma\gamma\tilde{G}\tilde{G}$, that is, two 
acoplanar photons plus missing energy, since the gravitinos stay undetected.
Searches for these kind of events have been unsuccesful and limits on $m_\chi$
around 72~GeV have been set within some particular models~\cite{ref:lopez}.
%
%
\section{Four-jet anomaly}
\label{sect:4-jet}
\begin{figure}
\begin{center}
\place{0.9}{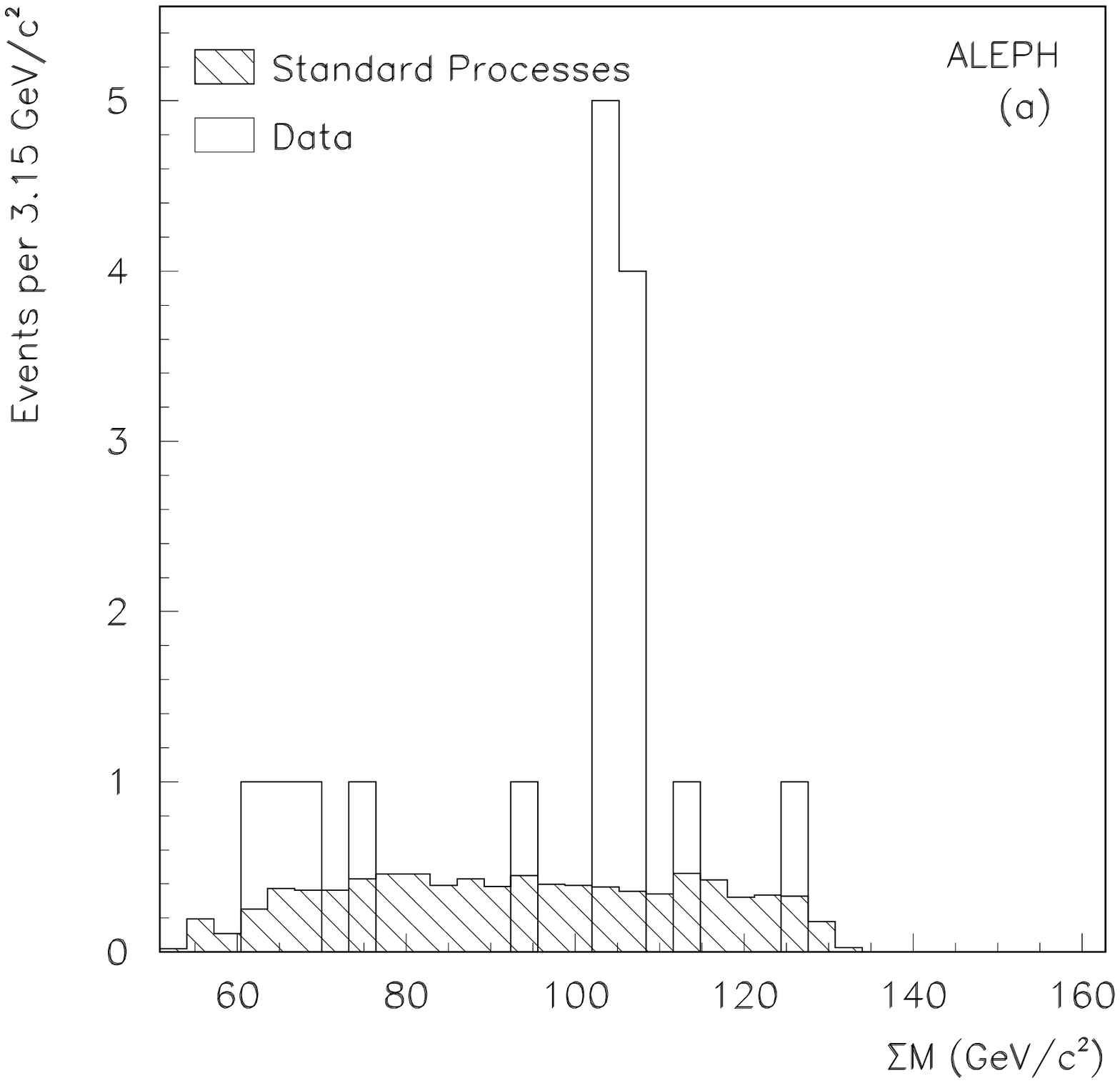}{fig:4j_130}
{Distribution of the sum of the invariant masses of the two jet pairs with 
smallest invariant mass difference in four-jet events, from ALEPH at 
130~GeV~\cite{ref:aleph_133}.}
\end{center}
\end{figure}
In the autumn 1995 run at energies close to 130~GeV, the ALEPH 
collaboration reported an excess
of four-jet events in which the sum of the two
jet-pair invariant masses that differed the least was peaked at about 
105~GeV, as shown in fig.~\ref{fig:4j_130} from
ref.~\cite{ref:aleph_133}. The excess was not confirmed by the other three LEP
collaborations.
%
\begin{figure}
\begin{center}
\place{0.9}{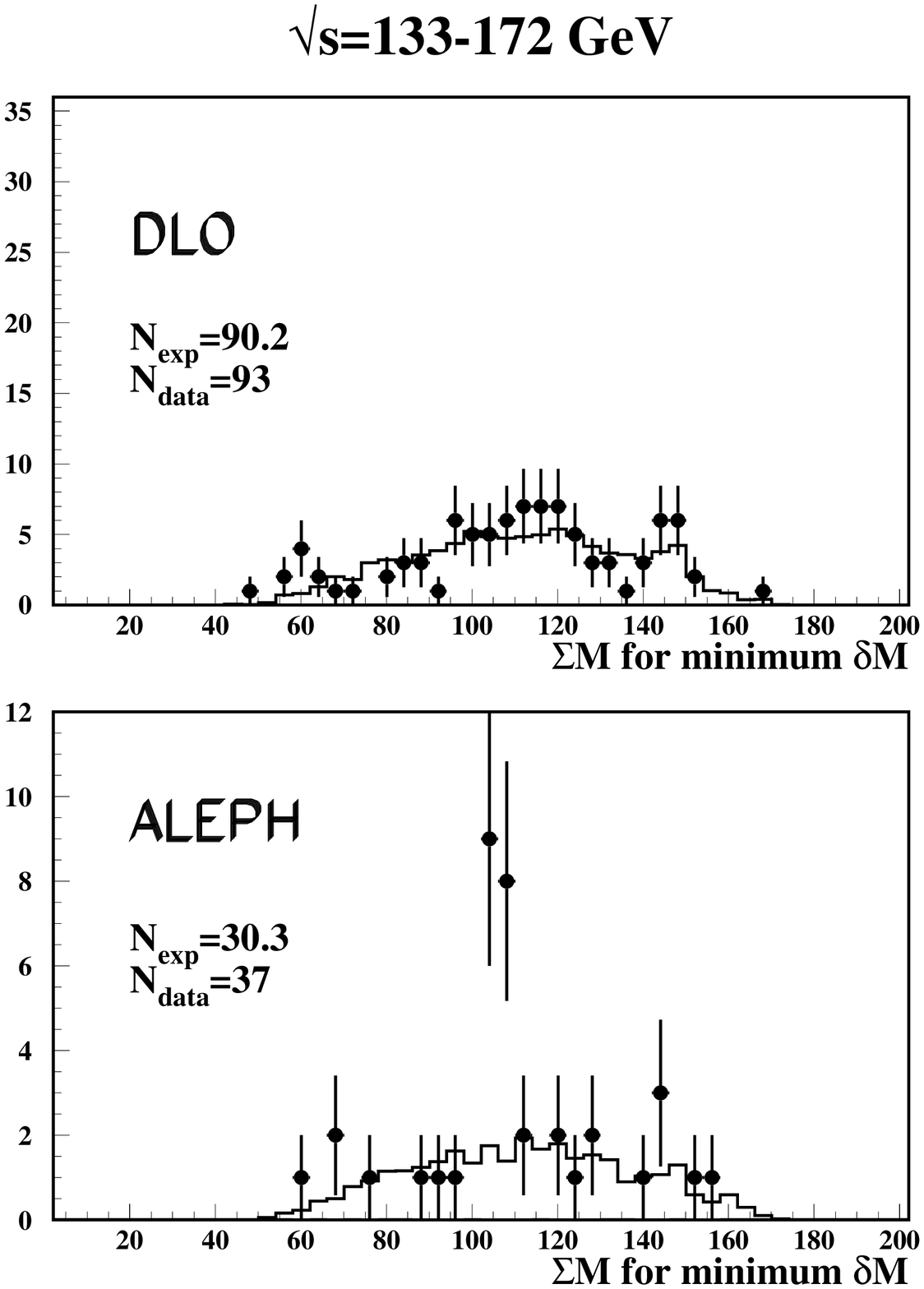}{fig:4j_all}
{Distribution of the sum of the invariant masses of the two jet pairs with 
smallest invariant mass difference in four-jet events at 130, 161 and 172~GeV
put together,
from DELPHI, L3 and OPAL (top plot) and from ALEPH (bottom plot).}
\end{center}
\end{figure}
Four-jet events have been studied by all LEP collaborations at both runs at
161~GeV and 172~GeV. While ALEPH seems to confirm their initial finding,
although with lower significance,
DELPHI, L3 and OPAL continue to see no
deviation from the Standard Model expectations. A LEP-wide
working group on the subject has concluded~\cite{ref:4jwg}
that the four experiments are able 
to select and reconstruct four-jet events with similar efficiencies and mass
resolutions. Figure~\ref{fig:4j_all} shows the combined results at all
energies from 130 to 172~GeV for DLO (that is, DELPHI, L3 and OPAL combined)
and for ALEPH. The discrepancy is clear.

The study concludes that the probability for the ALEPH observation
to be due to a statistical fluctuation is about $7\times 10^{-4}$, while the
probability for both the ALEPH and the DLO observations to be compatible with
a signal at 105~GeV is also about $7\times 10^{-4}$! It is clear that, so far,
the origin of the effect is not understood.
\section{Constraints on possible explanations of HERA events}
\label{sect:HERA}
If the excess of high $Q^2$ events seen in H1 and ZEUS at HERA~\cite{ref:HERA}
is due to new physics affecting the coupling between quarks and electrons, 
there could be also effects in the process $e^+e^-\to q \bar{q}$, very well
studied at LEP. The OPAL collaboration has investigated two
scenarios~\cite{ref:lq_opal}, either with the presence of four-fermion
contact iteractions between $e-e-q-q$, 
or with the exchange of a scalar particle
in the t-channel of the reaction $e^+e^-\to q \bar{q}$.

By measuring the cross section for hadron production and, separately, the
ratio of $b\bar{b}$ final states to all hadronic final states, both at 161 and
at 172~GeV, OPAL has put limits on the mass scale appearing in the contact-term
lagrangian between 1.0 and 2.5~TeV, depending on the type of interaction and on
whether the interaction affects one up-type quark or one down-type quark.
It should be noted that, in spite of the huge statistics accumulated at LEP1,
this kind of search is more sensitive at LEP2 because at the Z peak the 
interference between the new amplitude (purely real) and the dominant Z 
amplitude (almost purely imaginary) almost vanishes and one is left with the
purely new effect squared.

If a t-channel exchange of a scalar leptoquark or a scalar quark violating
R-parity is assumed, some constraints on possible models explaining the
HERA excess events can be obtained. The constraints will become truly severe
once the 1997 data is analysed.
\section{Summary}
\label{sect:summary}
After analysing the data from the first year of LEP2, no deviation from the
Standard Model predictions has been found:
\begin{itemize}
\item The direct measurement of the W mass agrees with the previous indirect
determinations using LEP1/SLD data and assuming the Standard Model.
\item Limits on Higgs and Supersymmetry particles have been greatly extended.
For example, the Standard Model Higgs has to be heavier than 71~GeV, the SUSY
Higgses $h$ and $A$ heavier than 62.5~GeV, and the
lightest chargino heavier than 85~GeV in most of the SUSY parameter space.
\item The four jet anomaly reported by ALEPH in 1995 remains a mistery, having
been confirmed by ALEPH in 1996 but not seen by any of the other LEP 
experiments.
\item Analysing the reaction $e^+e^-\to q \bar{q}$ the LEP experiments can 
start to get interesting constraints on models trying to explain the high $Q^2$
HERA events
\end{itemize}
\section*{Acknowledgments}
It has been a great pleasure to attend this conference in such a nice setting
and with such an attendance. I am most thankful to 
Prof.~Klapdor-Kleingrothaus
for his invitation to give this talk
and to him and all his team for making the whole week so
enjoyable. I would also like to thank Dr.~Michael Schmitt for his help in the
preparation of this talk.

\end{document}